\documentclass[final,5p,times,twocolumn]{elsarticle}

\usepackage{hyperref}
\hypersetup{
    colorlinks,
    linkcolor={red!50!black},
    citecolor={blue!50!black},
    urlcolor={blue!80!black}
}

\usepackage{amssymb}

\usepackage{amsmath}
\DeclareMathOperator{\tr}{tr}
\DeclareMathOperator{\C}{C}

\usepackage{tikz}
\usetikzlibrary{decorations.pathreplacing}
\definecolor{mygreen}{rgb}{0,0.5,0}
\definecolor{myblue}{rgb}{0,0,1}

\usepackage{mwe}

\usepackage{booktabs}
\usepackage{subcaption}
\usepackage{graphicx}

\journal{Physics Letters B}

\begin{document}


\begin{frontmatter}




\title{Hybrid Monte Carlo Simulation with Fourier Acceleration of
       the $N=2$ Principal Chiral Model in two Dimensions}

\author[label1]{Roger Horsley}
\affiliation[label1]{organization={University of Edinburgh},
             addressline={School of Physics and Astronomy},
             city={Edinburgh},
             postcode={EH9 3HD},
             country={UK}}

\author[label1]{Brian Pendleton}

\author[label1]{Julian Wack}




\begin{abstract}
   Motivated by the similarity to QCD, specifically the property of asymptotic freedom, we simulate the dynamics of the \mbox{SU(2) $\times$ SU(2)} model in two dimensions using the Hybrid Monte Carlo algorithm. By introducing Fourier Acceleration, we show that critical slowing down is largely avoided and increases the simulation efficiency by up to a factor of 300. This yields numerical predictions at a precision exceeding that of existing studies and allows us to verify the onset of asymptotic scaling.
\end{abstract}






\end{frontmatter}





\section{Introduction}
\label{sec:introduction}

Hybrid Monte Carlo (HMC) was first introduced in \cite{Duane1987HMC} and developed to allow simulations of quantum chromodynamics (QCD) including fermions. However, like other Monte Carlo methods, it suffers from critical slowing down as a critical point such as a second or higher order phase transition is approached. In general, the number of steps required to obtain a statistically independent configuration, the autocorrelation time $\tau$, scales as a power law of the correlation length $\xi$ or the lattice spacing $a$. Namely, 
\begin{equation}
    \tau \propto \xi^z \propto a^{-z}    
\label{eq:CSD}
\end{equation}
where $z$ quantifies the severity of slowing down and typically $z \sim 2$. Employing Fourier Acceleration (FA) causes the system's long and short wavelength modes to evolve at a more equal rate \cite{Duane1986_accelgauge}, therefore reducing the autocorrelation time significantly. Gauge invariant theories such as QCD mix the high and low frequency modes, causing the effect of FA to be masked. Nevertheless, FA has been applied successfully in SU(2) \cite{Duane1988_GiFA} and SU(3) \cite{Davies1990} gauge theory simulations. Present attempts to bypass this issue in the HMC context include a FA inspired version of Riemannian Manifold HMC \cite{Nguyen2021} and employing a softly gauge-fixed FA HMC algorithm \cite{Sheta2021}.\\ 
The 2-dimensional chiral model presents a semi-realistic context to illustrate the advantage of using FA in HMC as it is an asymptotically free (but non-gauge invariant) theory which develops massive states dynamically and is analytically understood \cite{Rossi1994, Rossi1994_2, Guha1984, Balog1992}. For the $N=2$ case, we show that FA achieves its goal: a significant reduction of $z$ compared to unaccelerated HMC.\\
The paper is organised as follows. Firstly, we introduce the HMC algorithm and its extension by FA. The chiral model is discussed in Section~\ref{sec:chiral_model}. We contrast numerical results of the internal energy density against analytical expansions before computing the $\beta$ function and verifying the onset of asymptotic scaling. In Section~\ref{sec:crit_slowdown}, the reduction in critical slowing is quantified.


\section{The Hybrid Monte Carlo Algorithm and Fourier Acceleration}
\label{sec:HMC+FA}

For a Euclidean action $S$, the HMC algorithm allows us to efficiently generate field configurations $\phi$ according to the target distribution $\exp(-S(\phi))$. This is achieved by constructing a Hamiltonian from the action and a kinetic term, featuring auxiliary momenta $p$ conjugate to the degrees of freedom of the theory. By solving Hamilton's equations, one obtains a candidate for the Markov chain which, based on the Metropolis update \cite{Metropolis1953}, has a high acceptance rate while exploring distant regions of the $(\phi, p)$ phase space.\\
We choose standard Gaussian momenta, such that the Hamiltonian reads
\begin{equation}
    H(\phi, p) = \frac{1}{2} p^T \cdot p + S(\phi)
\label{eq:Ham_original}
\end{equation}
where sums over lattice sites are implied by the dot notation.\\ 
We solve the equations of motion  (EOM) numerically using the leapfrog scheme for a fixed unit trajectory length. Subject to this constraint, and as HMC performs optimally with an acceptance rate $\mathcal{R}$ of $0.65$ \cite{Neal2012}, the number and size of integration steps are calibrated such that $\mathcal{R} \in [0.6, 0.75]$.

\subsection{Modified Dynamics}
\label{sec:mod_dyms}
As the lattice spacing $a$ is reduced, increasingly high-frequency modes are contained in the simulation, imposing a progressively strict upper bound on the integration step size. For a fixed number of integration steps, many sweeps are required to yield a notable change in the physical low-energy modes resulting in large autocorrelation times. One possibility to reduce the magnitude of this is to modify the dynamics. Denoting the action kernel by $K$, the new Hamiltonian becomes
\begin{equation}
    H(\phi, p) = \frac{1}{2} p^T \cdot K^{-1} \cdot p + S(\phi)
\label{eq:Ham_mod}
\end{equation}
such that the modified dynamics are governed by
\begin{equation}
    \dot{\phi} = K^{-1} \cdot p, \quad \dot{p} = - \frac{\partial S}{\partial \phi}.
\end{equation}
Comparing the former (relating the time evolution of the degree of freedom to the conjugate momenta) with the relation between position and momentum in classical mechanics suggests interpreting $K^{-1}$ as a mass. For a kernel involving derivatives, this mass is momentum-dependent such that the evolution of high momentum modes is slowed down while slowly evolving modes are accelerated. The overall more uniform evolution speed reduces the degree of critical slowing down.\\
While the inversion of $K$ in configuration space is possible in principle, the computational cost proliferates with the lattice size, motivating us to go to Fourier space where the kernel is diagonal and thus trivially inverted. Denoting the Fourier transform by a tilde, the EOM for the degrees of freedom used in practice is
\begin{equation}
    \dot{\phi} = \mathcal{F}^{-1} \left [ \tilde{K}^{-1} \cdot \tilde{p} \right].
\end{equation}

\subsection{Distribution of Momenta}
Inserting $K^{-1}$ in the kinetic term changes the distribution of the momenta and requires the sampling to be done in Fourier space by the aforementioned inversion-related problem. We will only consider real degrees of freedom such that for a $D$ dimensional cubic lattice of side length $L$, the momentum Fourier modes $p_\mathbf{k}$ are subject to Hermitian symmetry
\begin{equation}
    \tilde{p}_{k_1, \dots, k_D} = \tilde{p}^*_{L-k_1, \dots, L-k_D} \quad \text{with} \quad k_i = 1, \dots, L-1.
\end{equation}
Consequently, only $L^D$ of the $2L^D$ real and imaginary components are independent. Organising these into a real-valued object $\tilde{\Pi}_\mathbf{k}$ such that  $\sum_\mathbf{k} \vert \tilde{p}_\mathbf{k} \vert^2 = \sum_\mathbf{k} \tilde{\Pi}_\mathbf{k}^2$, results in drawing $\tilde{p}_\mathbf{k}$ from the distribution
\begin{align}
    \exp{ \left ( -\frac{1}{2} \tilde{p}^{\dagger} \cdot \tilde{K}^{-1} \cdot \tilde{p} \right )}
    &= \prod_\mathbf{k} \exp{ \left ( -\frac{1}{2L^D} \vert \tilde{p}_{\mathbf{k}} \vert^2 \tilde{K}^{-1}_\mathbf{k} \right )} \nonumber \\ 
    &= \prod_\mathbf{k} \exp \left ( -\frac{1}{2L^D} \tilde{\Pi}_\mathbf{k}^2 \tilde{K}^{-1}_\mathbf{k} \right).
\label{eq:mom_dist}
\end{align}  
Hence, the components $\tilde{\Pi}_\mathbf{k}$ are sampled from independent Gaussians with zero mean and standard deviation $\sqrt{L^D / \tilde{K}^{-1}_\mathbf{k}}$. Reconstructing $\tilde{p}$ from $\tilde{\Pi}$ and taking the inverse Fourier transform yields the required real space momentum samples, following the modified kinetic term of Equation~\ref{eq:Ham_mod}. In~\ref{app:momentum_dist} we describe the construction of $\tilde{\Pi}_\mathbf{k}$ explicitly for the $D=2$ case.


\section{The SU(2) $\times$ SU(2) Chiral Model}
\label{sec:chiral_model}

Let $\sigma_i$ be the Pauli matrices such that the SU(2) matrix-valued field $\phi(x)$ in the fundamental representation is parameterised by $\pmb{\alpha}(x) \in \mathbb{R}^3$ and can be expressed as 
\begin{align}
    \text{SU}(2) \ni \phi = e^{i \alpha_i \sigma_i} &= \cos \alpha \pmb{1}_2 + i \sin \alpha \left ( \hat{\pmb{\alpha}} \cdot \pmb{\sigma} \right ) \nonumber \\
    &= c_0 \pmb{1}_2 + i c_i \sigma_i 
\label{eq:phi_parameterisation}
\end{align}
with $\alpha = |\pmb{\alpha}|, ~\hat{\pmb{\alpha}} = \pmb{\alpha} / \alpha$ and where we defined the parameter vector $c_{\mu} = (c_0, c_i)$ for $c_0=\cos \alpha,~c_i=\hat{\pmb{\alpha}}_i \sin \alpha$. The unitarity constraint and the $\det \phi = 1$ requirement impose the condition
\begin{equation}
    |c|^2 = c_0^2 + c_1^2 + c_2^2 + c_3^2 = 1,
\label{eq:norm_a}
\end{equation}
allowing us to interpret the $4$ real scalar fields $c_\mu$ as the degrees of freedom constraint to the $S^3$ group manifold.

For a lattice with sites $x$ and with basis vectors $\mu$ (both being $D=2$ dimensional vectors), we choose the action of the principal chiral model as
\begin{align}
    S(\phi) &= - \frac{\beta}{2} \sum_{x,\mu} \tr \left \{ \phi_x^{\dagger} \phi_{x+\mu} + \text{h.c.} \right \} \nonumber \\
    &= -\beta \sum_{x,\mu} \Re \tr \left \{ \phi_x^{\dagger} \phi_{x+\mu} \right \} 
    = -\beta \sum_{x,y} \: \Re \tr \left \{ \phi_x^{\dagger} K_{x,y} \phi_y \right \}
\label{eq:action}
\end{align}
where $\beta$ is defined in terms of the coupling constant $g_0$ via \mbox{$\beta = 4/(N g_0^2)$}. The second line motivates us to choose the action kernel as 
\begin{equation}
    K_{x,y} = \sum_{\mu} \left ( \delta_{x,y+\mu} + \delta_{x,y-\mu} - 2\delta_{x,y} \right ).
\label{eq:kernel}
\end{equation}

\subsection{Equations of Motion}
To increase readability, we largely suppress the dependence on the Hamiltonian time $t$, meaning $\phi_x \equiv \phi(x,t)$, but for clarity write the discrete EOM using the more explicit notation.\\
From the parameterisation in Equation~\ref{eq:phi_parameterisation}, the momenta \mbox{$p_x \equiv p(x,t)$} conjugate to the degrees of freedom $\alpha_i(x,t)$ may be defined through the left time derivative of $\phi_x$:
\begin{align}
    \frac{d}{dt} \phi_x &= \dot{\phi}_x = i \dot{\alpha}_i(x,t) \sigma_i \phi_x = i p_x \phi_x \label{eq:phi_eom} \\
    p_x &= p_i(x,t) \sigma_i \equiv \dot{\alpha}_i(x,t) \sigma_i \in \mathfrak{su}(2).
\end{align}
For the leapfrog integrator, the momenta are defined on a time lattice shifted by half a step size $\epsilon$ relative to the field. The discrete analogue of Equation~\ref{eq:phi_eom} is 
\begin{equation}
    \phi(x, t+\epsilon) - \phi(x,t) = i p \left ( x,t + \frac{\epsilon }{2} \right ) \phi(x,t) \epsilon
\end{equation}
which, to linear order, can be written as the exponential update
\begin{equation}
    \phi(x, t+\epsilon) = \exp \left \{ i p \left ( x,t + \frac{\epsilon }{2} \right ) \epsilon  \right \} \phi(x,t) + \mathcal{O} \left( \epsilon^2 \right) \label{eq:discret_phi_EoM}.
\end{equation}
The $N=2$ model allows us to evaluate the exponential exactly such that the dynamics remain on the group manifold. The finite machine precision introduces a small error which requires us to re-enforce Equation~\ref{eq:norm_a} occasionally. For our simulations, we found that doing so every $10^4$ trajectories is sufficient.\\
For the Hybrid algorithm, we introduce a Gaussian kinetic term 
\begin{align}
    T = \frac{1}{4} \sum_x \tr p_x p_x =  \frac{1}{2} \sum_{x,i} p_i(x,t)p_i(x,t)
\end{align}
and deduce the EOM for the momenta by imposing energy conservation. Namely
\begin{equation}
    \frac{d}{dt} H = \dot{T} + \dot{S} = 0 \quad \text{with} \quad \dot{T} = \frac{1}{2} \sum_x \tr p_x \dot{p}_x
\label{eq:conservation_H}
\end{equation}
and, upon using Equations~\ref{eq:action} and~\ref{eq:phi_eom},
\begin{equation}
    \dot{S} = -i\frac{\beta}{2} \sum_{x,\mu} \tr \left \{ p_x \left ( \phi_x \phi^{\dagger}_{x+\mu} - \phi_{x+\mu}\phi^{\dagger}_x \right ) + p_{x+\mu} \left ( \phi_{x+\mu} \phi^{\dagger}_x - \phi_x\phi^{\dagger}_{x+\mu} \right ) \right \}.
\end{equation}
Relabelling $x \to x + \mu$ in the second term yields
\begin{align}
    \dot{S} &=  -i\frac{\beta}{2} \sum_{x,\mu} \tr \left \{ p_x \left ( \phi_x \left ( \phi^{\dagger}_{x+\mu} + \phi^{\dagger}_{x-\mu} \right ) - \text{h.c.} \right ) \right \} \nonumber \\
    &= i \frac{\beta}{2} \sum_{x,\mu} \tr \left \{ p_x \left ( \left ( \phi_{x+\mu} + \phi_{x-\mu} \right ) \phi^{\dagger}_x - \text{h.c.} \right ) \right \}.
\end{align}
Based on Equation~\ref{eq:conservation_H}, we thus identify
\begin{equation}
    \dot{p}_x = -i \beta \sum_{\mu} \left ( \left ( \phi_{x+\mu} + \phi_{x-\mu} \right ) \phi^{\dagger}_x - \text{h.c.} \right )
\label{eq:pi_eom}
\end{equation}
such that the momenta are updated according to 
\begin{equation}
    p \left ( x, t+\frac{\epsilon }{2} \right ) = p \left ( x,t-\frac{\epsilon }{2} \right ) + \dot{p}(x,t) \epsilon.
    \label{eq:pi_eom_discrete}
\end{equation}

\subsection{Fourier Acceleration}
Using Equation~\ref{eq:kernel} and the Fourier representation of the Dirac delta, one quickly finds
\begin{equation}
    \tilde{K}_{k,k'} = \sum_{\mu} 4 \sin^2 \left ( \pi \frac{k_{\mu}}{L} \right ) \delta_{k,k'} \quad \text{with} \quad k_{\mu} \equiv k \cdot \mu.
\end{equation}
To assure that the inverse is defined for all $k$, the acceleration mass parameter $M$ is introduced, yielding
\begin{equation}
    \tilde{K}^{-1}_{k,k'}(M) = \frac{\delta_{k,k'}}{\sum_{\mu} 4 \sin^2 \pi \frac{k_{\mu}}{L} + M^2}.
\end{equation} 
For free theories, taking $M$ as the physical mass fully avoids critical slowing down. Considering an asymptotically free theory, where the modes decouple as the coupling tends to zero, motivates choosing this kernel. While the optimal value of $M$ for interacting theories is less clear \cite{Duane1986_accelgauge}, the physics is independent of $M$, allowing us to fix $M=1/10$ until Section~\ref{sec:crit_slowdown}. There we show that, provided $1/M$ is of the same order as $\xi$ measured in units of the lattice spacing, the simulation efficiency only weakly depends on the precise value of $M$.\\
For a two-dimensional lattice, the modified Hamiltonian used in practice is therefore given by
\begin{equation}
    H(\phi, p) = \frac{1}{2L^2} \frac{\sum_i |\tilde{p}_i(k)|^2}{4 \sin^2 \pi \frac{k_1}{L} + 4 \sin^2 \pi \frac{k_2}{L} + M^2} + S(\phi).
\end{equation}
Based on the general discussion in Section~\ref{sec:mod_dyms}, modifying the dynamics changes the EOM for the degrees of freedom to 
\begin{equation}
    \dot{\alpha}_i(x,t) = K^{-1}(M) p_i(x,t).
\label{eq:alpha_dot_modified}
\end{equation}
The field is then updated according to 
\begin{equation}
    \phi(x, t+\epsilon) = \exp \left \{ i \mathcal{F}^{-1} \left [ \tilde{K}^{-1}(M) \tilde{p} \left (k,t + \frac{\epsilon}{2} \right ) \right ] \epsilon  \right \} \phi(x,t).
\end{equation}

\subsection{Strong and Weak Coupling Expansions}
Considering a $D$-dimensional lattice of volume $V$ and free energy density $f$ for a general SU($N$) $\times$ SU($N$) model, we define the internal energy $e$ as
\begin{align}
    e = \frac{N}{D} \frac{\partial f}{\partial \beta} \quad \text{with} \quad f = \frac{1}{V N^2} \log Z.  
\label{eq:e_def}
\end{align}
\pagebreak 
Through the standard observation of statistical physics \mbox{$\frac{\partial }{\partial \beta} \log Z = \left \langle -\frac{S}{\beta}\right \rangle$} and specialising to $N=2, D=2$ we obtain
\begin{equation}
    e = \frac{1}{4V} \left \langle -\frac{S}{\beta}\right \rangle.
\label{eq:internal_energy}
\end{equation}
Expansions for $e$ in powers of the coupling for the SU($N$) $\times$ SU($N$) principal chiral model are discussed in detail in the literature with the results summarised in \cite{Rossi1994, Brihaye1984, Green1981}. After accounting for a relative factor of 4 in the definition of $\beta$, we extract the strong coupling expansion
\begin{equation}
    e = \frac{1}{2} \beta + \frac{1}{6} \beta^3 + \frac{1}{6} \beta^5 + \mathcal{O} \left( \beta^7 \right)
\label{eq:strong_coupling}
\end{equation}
and the weak coupling expansion
\begin{equation}
    e = 1 - \frac{3}{8\beta} \left (1 + \frac{1}{16 \beta} + \left ( \frac{1}{64} + \frac{3}{16}Q_1 + \frac{1}{8}Q_2 \right) \frac{1}{\beta^2}  \right ) + \mathcal{O} \left( \beta^{-4} \right).
\label{eq:weak_coupling}
\end{equation}
The latter is obtained by computing loop corrections to the mean field result, resulting in the numerical constants \mbox{$Q_1 = 0.0958876$ and $Q_2 = -0.067$}.

\begin{figure}
    \centering
    \includegraphics[width=0.475\textwidth]{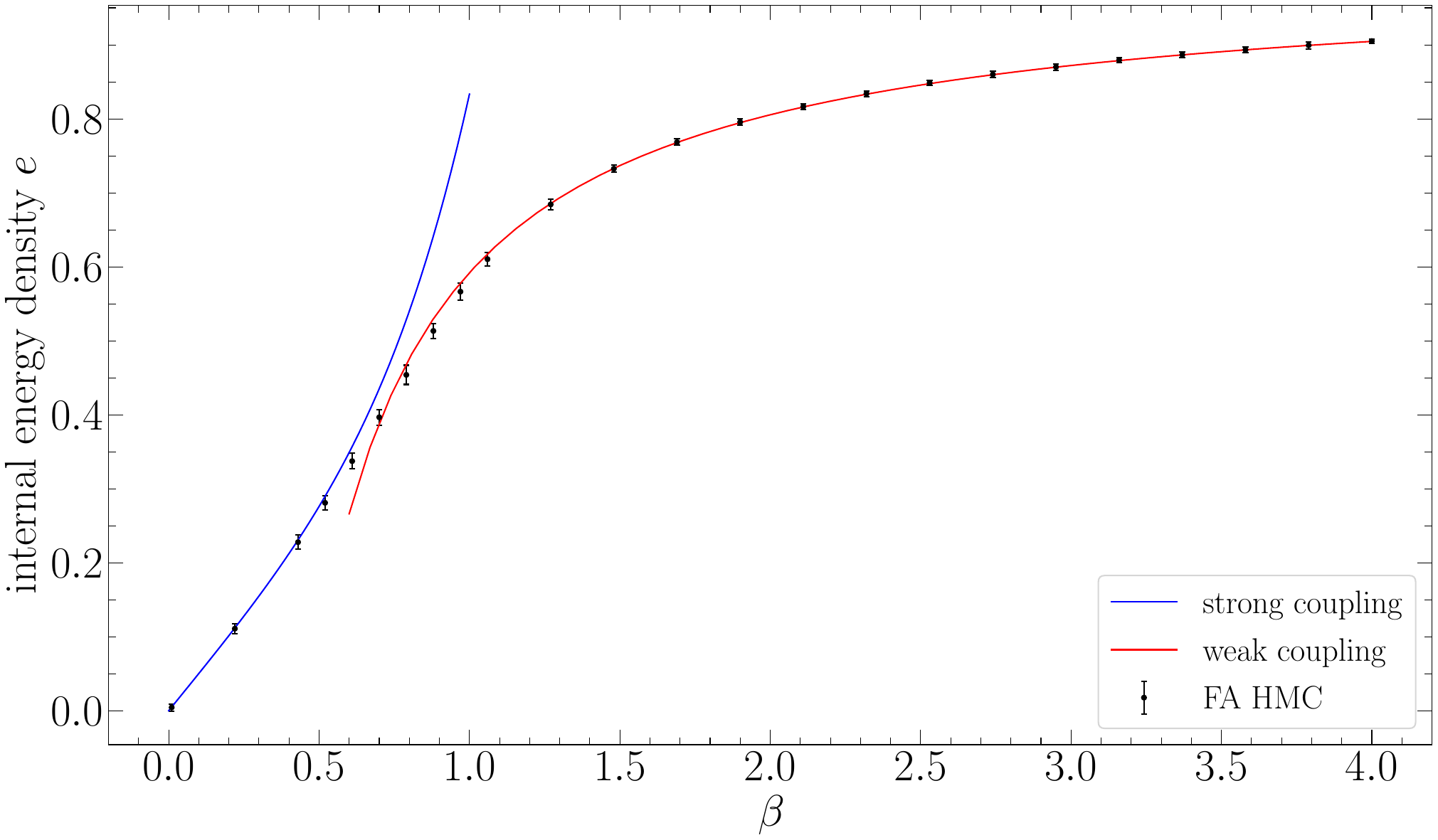}
    \caption{The internal energy density against $\beta$ based on simulating $5000$ trajectories ($10\%$ burn in and $L=16$) per data point using Fourier accelerated HMC with a fixed acceleration mass of $M=1/10$. Error bars are enlarged by a factor of 10. The strong (blue) and weak (red) coupling expansions are superimposed.}
    \label{fig:coupling_exp}
\end{figure}

The numerical result of Equation~\ref{eq:internal_energy} and the analytic predictions of Equations~\ref{eq:strong_coupling},~\ref{eq:weak_coupling} are plotted in Figure~\ref{fig:coupling_exp} against different values of $\beta$ for $L=16$.
Simulating $5000$ trajectories for each data point was sufficient to obtain numerical results whose precision exceeds that of existing studies which use a local updating algorithm \cite{Guha1984, Guha1984_2}. To account for autocorrelations, we quote the standard error on the mean, corrected by the square root of the integrated autocorrelation time $\tau$, as the statistical uncertainty $\delta$ of our measurements. Specifically, for $\mathcal{M}$ observations,
\begin{equation}
    \delta = \sqrt{\tau} \frac{\sigma}{\sqrt{\mathcal{M}}} \quad \text{where} \quad \tau = 1 + 2 \sum_{t=1}^{K} \rho(t)
\label{eq:IAT_corrected_SEM}
\end{equation}
and $\rho(t)$ denotes the autocorrelation function while the cut $K$ is chosen following \cite{Caracciolo1986}. The error bars in Figure~\ref{fig:coupling_exp} have been enhanced by a factor of 10 for better readability. In their region of applicability, the analytic predictions are well matched by the numerical results. However, in the range $\beta \in [0.5, 1.0]$ the expansions lose applicability and the lattice simulation offers corrections of up to $6\%$.

\subsection{Correlation Function}
\begin{figure}
    \centering
    \scalebox{1.15}{
            \begin{tikzpicture}
                \draw[step=1, gray, very thin] (-0.2,-0.2) grid (4.2,2.2);
                \draw[myblue, thick] (0,0) -- (0,2);
                \draw[myblue, thick] (3,0) -- (3,2);
    
                \coordinate (F1) at (0,2);
                \coordinate (F2) at (3,1);
    
                \node [circle, label=45:{$(v,i)$}, fill=black, scale=0.3] at (F1){};
                \node [circle, label=45:{$(w,i+d)$}, fill=black, scale=0.3] at (F2){};
                
                \draw[->, mygreen, thick, opacity=0.5] (0,2) -- (3,2);
                \draw[->, mygreen, thick] (0,2) -- (3,1);
                \draw[->, mygreen, thick, opacity=0.5] (0,2) -- (3,0);
    
                \draw [decorate,decoration={brace,amplitude=5pt,mirror,raise=0ex}] (1,0) -- (2,0) node[midway,yshift=-1em]{$a$};
            \end{tikzpicture}
            }
        \caption{Composition of the wall-to-wall correlation function at separation $d=3$ in terms of point-to-point correlations.}
        \label{fig:wallwall_cor}
    \end{figure}
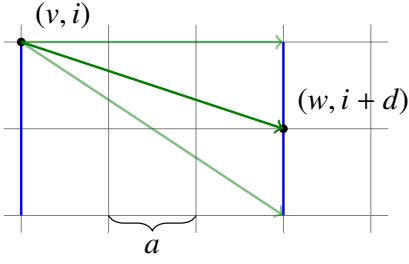

\begin{figure}
    \centering
    \includegraphics[width=0.475\textwidth]{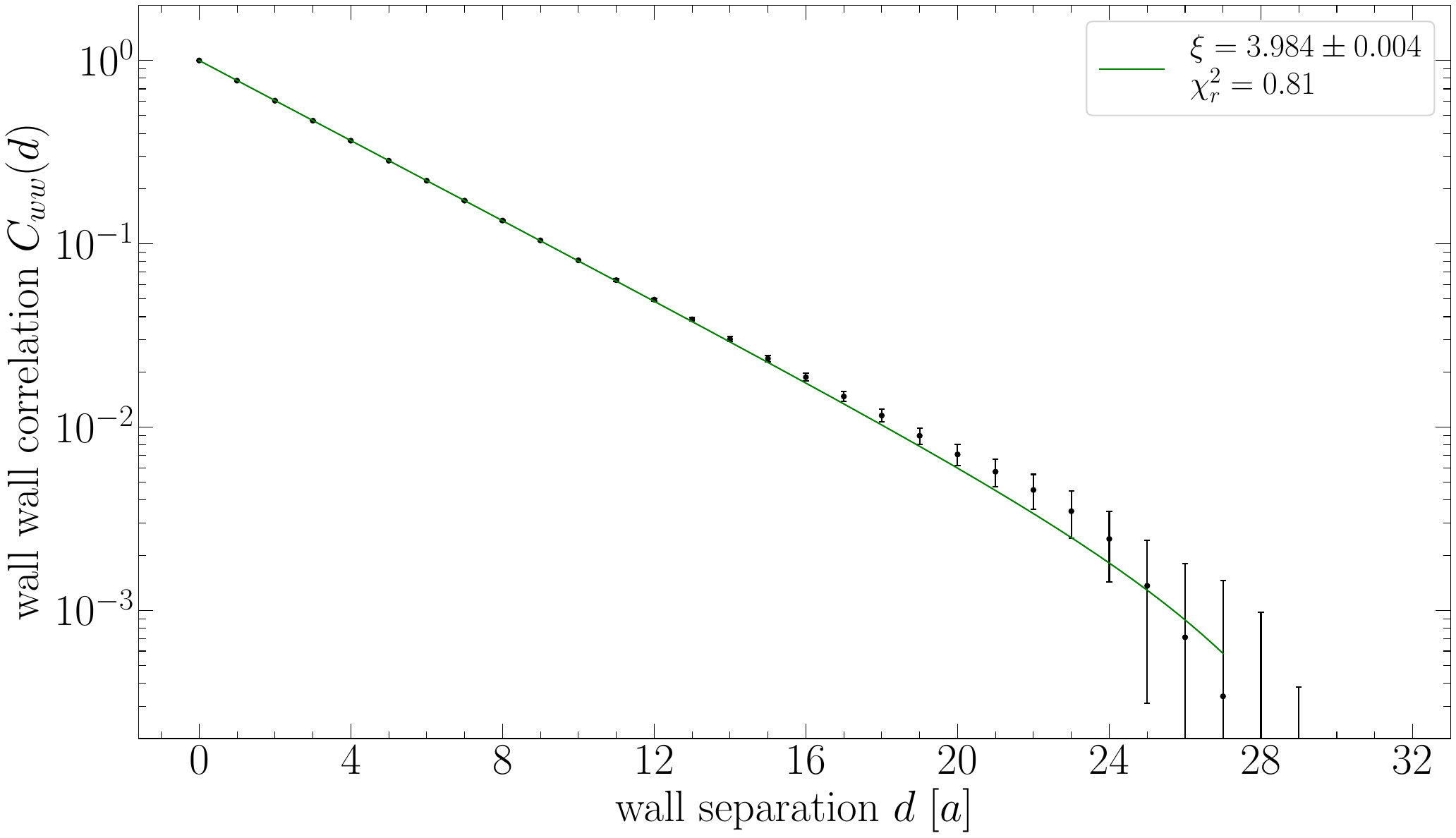}
    \caption{The wall-to-wall correlation function inferred from $10^5$ trajectories ($5\%$ burn in) at $\beta = 0.8667$ on a $L=64$ lattice and plotted on a logarithmic $y$ scale. A $\cosh$ function is fitted with the inferred correlation length and the associated $\chi^2$ per degree of freedom presented in the legend.}
    \label{fig:wallwall_plot}
\end{figure}

Due to the taken periodic boundary conditions, the two-point correlation function of $\phi$ with correlation length $\xi$ is symmetric and given by
\begin{equation}
    \C(d) \sim \cosh \left( \frac{d-L/2}{\xi} \right).
\label{eq:analytic_cor}
\end{equation}
To obtain $\xi$, we exploit this symmetry by fitting $C(d)$ to the average correlation function data at $d$ and $L-d$ on $d\in[0,L/2]$, effectively doubling the statistics.\\
We compute the correlation function based on wall-to-wall correlations which are defined as the average point-to-point correlations contained by two walls in the lattice with separation $d$. This is schematically shown in Figure~\ref{fig:wallwall_cor} and has the benefit that the number of wall pairs $P$ in each field configuration is large, increasing the statistics for $\C(d)$.
We omit the Hamiltonian time dependence and expand the lattice site $x$ into its coordinates $(v,i)$. The $m$th field configuration evaluated at the lattice site $(v,i)$ is thus denoted by $\phi_m(v,i)$. At a fixed value of $i$, averaging over the row indices $v$ and $w$ of the point pairs in Figure~\ref{fig:wallwall_cor} yields one measurement of the wall-to-wall correlation at separation $d$ for the currently considered configuration. Denoting the ensemble average as $\langle \dots \rangle_m$, the wall-to-wall correlation becomes
\begin{align}
    \C_{ww}(d) = \left \langle \frac{1}{P} \sum_i \left \{ \frac{1}{L^2} \sum_{v,w} 2 \Re \left ( \tr \phi_m(v,i) \phi^{\dagger}_m(w,i+d) \right ) \right \} \right \rangle_m.
\label{eq:wallwallcor}
\end{align}
The computation of the correlation function becomes expensive for large lattice sizes, motivating us to employ an efficient approach using Fourier transforms described in~\ref{app:cor_via_FT} and based on the close relationship between correlations and convolutions. Additionally, we normalise the correlation to its value at $d=0$ such that any constants in Equation~\ref{eq:wallwallcor} may be ignored. Again, we compute the error of the correlation function as the autocorrelation time corrected standard error on the mean. Figure~\ref{fig:wallwall_plot} shows the numerical results for $\beta = 0.8667$ on a $L=64$ lattice, the fitted Equation~\ref{eq:analytic_cor} with the associated reduced $\chi^2$, and the inferred correlation length.


\section{Asymptotic Scaling and the $\beta$ Function}

Asymptotic scaling requires the dimensionless ratio of any dimensionful quantity to the appropriate power of the regularisation scale $\Lambda$ to approach a constant as the coupling strength tends to zero. 
The $\beta$ function
\begin{equation}
    \beta(g_0^2) \equiv - a \frac{d}{d a} g_0^2 = -b_0 \left( g_0^2 \right) ^2 - b_1 \left( g_0^2 \right)^3 - b_2^L \left( g_0^2 \right)^4 - \dots
\label{eq:beta_func}
\end{equation}
is known at three-loop accuracy using lattice regularisation with
\begin{align}
    b_0 &= \frac{N}{8 \pi}, \quad b_1 = \frac{N^2}{128 \pi^2} \quad \text{and} \nonumber \\
    b_2^L &= \frac{1}{(2 \pi)^3} \frac{N^3}{128} \left ( 1 + \pi \frac{N^2-2}{2N^2} - \pi^2 \left (\frac{2N^4-13N^2+18}{6N^4} + 4G_1\right) \right )
\end{align}
where $G_1 = 0.04616363$ \cite{Rossi1994, Rossi1994_2}. Integrating Equation~\ref{eq:beta_func} and defining the inverse correlation length $\xi$, measured in units of the lattice spacing, as the mass $m$ yields
\begin{align}
    \frac{m}{\Lambda_L} &= \frac{1}{\xi a \Lambda_{L}} \nonumber \\
    &= \frac{1}{\xi} \exp{\left(\frac{1}{b_0 g_0^2}\right)} \left( b_0 g_0^2 \right)^{\frac{b_1}{b_0^2}} \exp{\left\{ \int_0^{g_0^2} du \frac{1}{\beta(u)} + \frac{1}{b_0 u^2} - \frac{b_1}{b_0^2 u} \right\}} \nonumber \\
    &= \frac{1}{\xi} \frac{\exp{(2\pi \beta)}}{\sqrt{2\pi \beta}} \exp{\left\{ \int_0^{\frac{4}{N \beta}} du \frac{1}{\beta(u)} + \frac{4\pi}{u^2} - \frac{1}{2 u} \right\}}.
\label{eq:mass_lambda_L}
\end{align}
The remaining part of this section is dedicated to testing the stability of Equation~\ref{eq:mass_lambda_L} with $\beta$ for the $N=2$ case against the mass gap scaling prediction
\begin{equation}
    \frac{m}{\Lambda_{L}} = \frac{32}{\sqrt{\pi e}} e^{\frac{\pi}{4}}
\label{eq:mass_lambda_cts}
\end{equation}
computed in \cite{Balog1992}. In order to do so, we evaluate the integral in Equation~\ref{eq:mass_lambda_L} numerically using the three-loop $\beta$ function and compute the correlation length $\xi$ for various values of $\beta$ based on $10^5$ trajectories (rejecting the first 2000 as burn in) using FA HMC with a fixed acceleration mass of $1/10$. The results are presented in Table~\ref{tab:asym_scaling}. For each value of $\beta$, the lattice size $L$ was chosen such that $L/\xi \gtrsim 10$ to reduce systematic errors due to finite size effects as suggested by \cite{Rossi1994}. The fitting range for Equation~\ref{eq:analytic_cor} was determined manually to yield $\chi^2_r \approx 1$ while excluding the noise-dominated regions of the correlation function data. Towards the lower end of the surveyed range of $\beta$, the data becomes quickly dominated by noise, such that only a small number of data points are suitable for the fitting, resulting in $\chi^2_r$ smaller than $1$.

\tabcolsep=2.1pt
\begin{table}
    \centering
    \footnotesize
    \begin{tabular*} {\linewidth}{@{\extracolsep{\fill}} c|c c c c c c c}
        \toprule
        $\beta$ & $0.6$ & $0.6667$ & $0.7333$ & $0.8$ & $0.8667$ & $0.9333$ & $1.0$ \\
        $L$ & $40$ & $40$ & $64$ & $64$ & $64$ & $96$ & $96$ \\
        \midrule
        $\xi$ & $1.421(2)$ & $1.781(2)$ & $2.268(2)$ & $2.950(3)$ & $3.984(4)$ & $5.542(6)$ & $7.93(1)$ \\
        $\chi^2_r$ & $0.20$ & $0.27$ & $0.39$ & $0.24$ & $0.81$ & $0.77$ & $1.02$ \\
        \bottomrule
        \multicolumn{8}{c}{} \\ 
        \toprule
        $\beta$ & $1.0667$ & $1.1333$ & $1.2$ & $1.2667$ & $1.3333$ & $1.4$ \\
        $L$ & $160$ & $160$ & $224$ & $400$ & $512$ & $700$ \\
        \midrule
        $\xi$ & $11.41(1)$ & $16.93(2)$ & $25.41(3)$ & $39.02(7)$ & $60.9(2)$ & $91.3(7)$ \\
        $\chi^2_r$ & $0.67$ & $0.84$ & $0.98$ & $1.05$ & $1.06$ & $0.84$ \\ 
        \bottomrule
    \end{tabular*}    
    \caption{The fitted correlation length, its error, and the associated $\chi_r^2$ for different pairings of $\beta$ and $L$. The lattice size was chosen conservatively to minimise finite size effects. All simulations employ the FA HMC algorithm with the acceleration mass fixed to $1/10$ and run for $10^5$ trajectories, rejecting the first 2000 as burn in.}
    \label{tab:asym_scaling}
\end{table}

\begin{figure}
    \centering
    \includegraphics[width=0.475\textwidth]{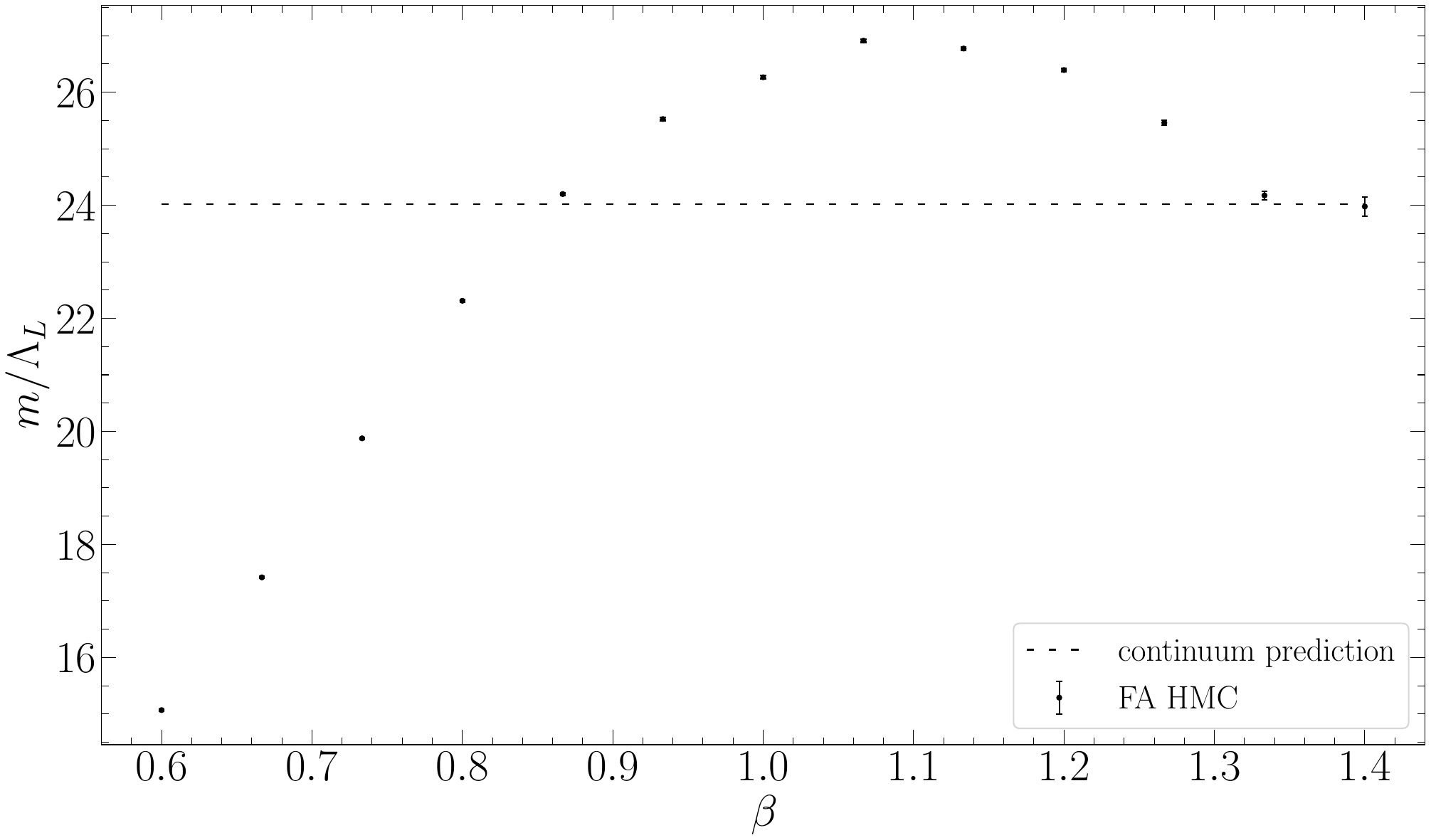}
    \caption{The mass over $\Lambda$ ratio of Equation~\ref{eq:mass_lambda_L} against $\beta$. Each data point is the result of fitting the correlation length to the averaged correlation function data based on $10^5$ trajectories (2000 burn in) using the lattice size $L$ specified in Table.~\ref{tab:asym_scaling}. The continuum prediction of Equation~\ref{eq:mass_lambda_cts} is shown as a dashed line.}
    \label{fig:asym_scaling}
\end{figure}

The collected data of Equation~\ref{eq:mass_lambda_L} is plotted against $\beta$ in Figure~\ref{fig:asym_scaling} with the continuum limit prediction of Equation~\ref{eq:mass_lambda_cts} indicated by the dashed line. The overall convergence to this constant is fairly poor, overshooting it by up to $12\%$ at intermediate values of $\beta$. This is consistent with the results of the $N=3,6,9,15$ cases and believed to be caused by a dip in the lattice beta function \cite{Rossi1994, Rossi1994_2}. The behaviour near the upper bound of the considered $\beta$ range suggests the onset of scaling.

\section{Critical Slowing Down Elimination}
\label{sec:crit_slowdown}
As suggested in \cite{Dagotto1987}, using the susceptibility $\chi$ as the observable to measure the effect of Fourier Acceleration is beneficial as it is dominated by low momentum modes and thus particularly prone to critical slowing down. Using any other variable should, however, result in the same conclusions but possibly with a less dramatic reduction in $z$. For a lattice of volume $L^2$, the susceptibility of a single configuration is defined as
\begin{equation}
    \chi = \frac{1}{L^2} \sum_{x,y} \Re \tr \phi_x \phi_y
\label{eq:def_chi}
\end{equation}
and can be viewed as the average point-to-point correlation of that configuration. In practice, performing the double sum over lattice sites quickly becomes the bottleneck for the analysis as $L$ gets large. It is therefore vital to compute $\chi$ via FFTs and the computationally much more efficient form given in Equation~\ref{eq:chi_crosscor}.

\subsection{Calibrating the Acceleration Mass}
\begin{figure}
    \centering
    \includegraphics[width=0.475\textwidth]{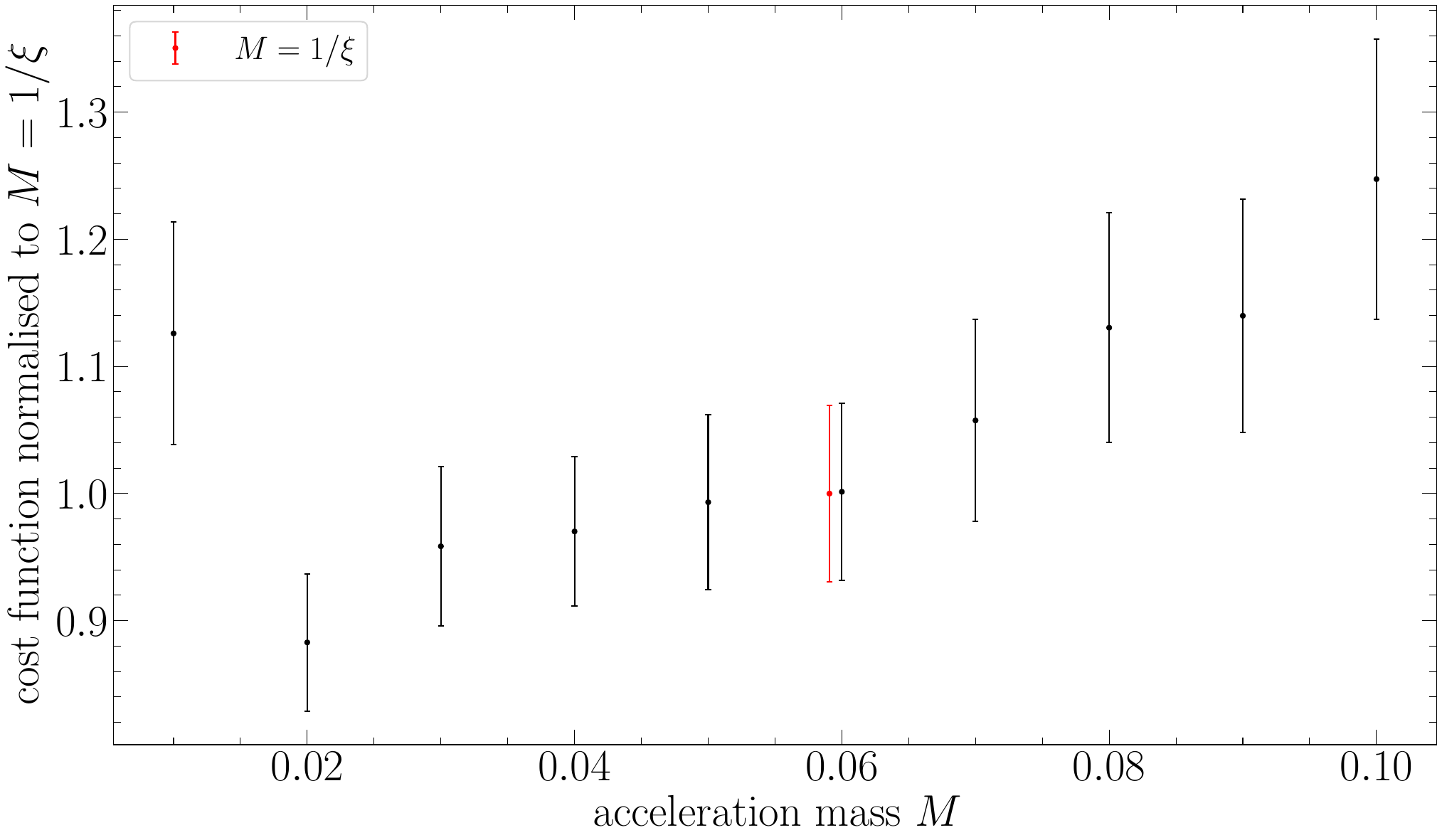}
    \caption{The cost function of Equation~\ref{eq:cost_func} for different values of the acceleration mass, normalised to the $M = 1/\xi$ choice. Each data point corresponds to $1.5 \times 10^4$ trajectories (with 10\% burn in) for $L = 160,\: \beta = 1.1333$.}
    \label{fig:accel_mass}
\end{figure}

To optimise the performance of FA HMC through the choice of $M$, we conducted a grid search at fixed \mbox{$\beta=1.1333$} based on the cost function
\begin{equation}
    \text{cost}(M) = \frac{\text{simulation time}}{\text{acceptance rate}} \sqrt{\tau}
\label{eq:cost_func}
\end{equation}
as the measurement error of an observable scales as $\sqrt{\tau}$. Figure~\ref{fig:accel_mass} details our conclusion that the conventional choice \mbox{$M = 1/\xi$} yields close to optimal acceleration, being roughly $10\%$ slower than the extremum of the scanned values of $M$. We adopt this choice henceforth and show in the following that this still offers a substantial speed-up compared to unaccelerated HMC.

\subsection{Results}
\begin{figure}[t]
    \centering
    \includegraphics[width=0.475\textwidth]{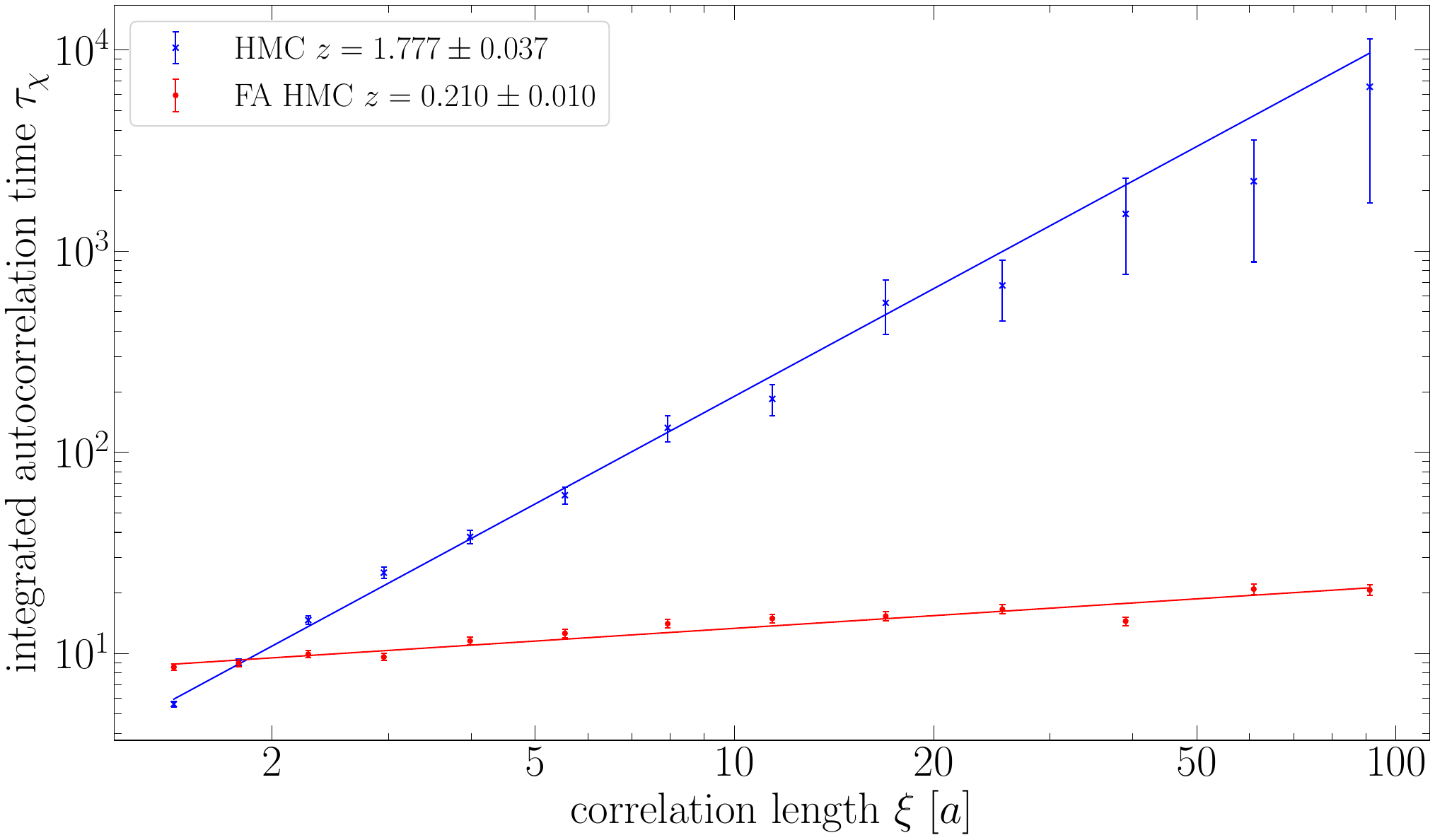}
    \caption{The integrated autocorrelation time of the susceptibility against the system correlation length $\xi$ on log-log axes for standard HMC (blue) and FA HMC (red) simulations. Fitted power laws are superimposed with the value of the critical exponent presented in the legend. Each data point corresponds to $10^5$ trajectories (5000 burn in) with values of $L$ and $\beta$ as specified in Table~\ref{tab:asym_scaling}.}
    \label{fig:slowdown}
\end{figure}

For each value pair $\beta, L$ from Table~\ref{tab:asym_scaling} we simulated $10^5$ trajectories (5000 burn in) using both standard HMC and FA HMC to deduce the susceptibility integrated autocorrelation time $\tau_\chi$ in either case. Due to terminating the sum in Equation~\ref{eq:IAT_corrected_SEM}, the error of $\tau_\chi$ scales as $\sqrt{\tau_\chi/\mathcal{M}}$, such that for a fixed number of observations $\mathcal{M}$, the uncertainty in $\tau_\chi$ increases towards the continuum. The data, as well as the fitted power law of Equation~\ref{eq:CSD}, are plotted in Figure~\ref{fig:slowdown}. 

\begin{figure}[t]
    \centering
    \includegraphics[width=0.475\textwidth]{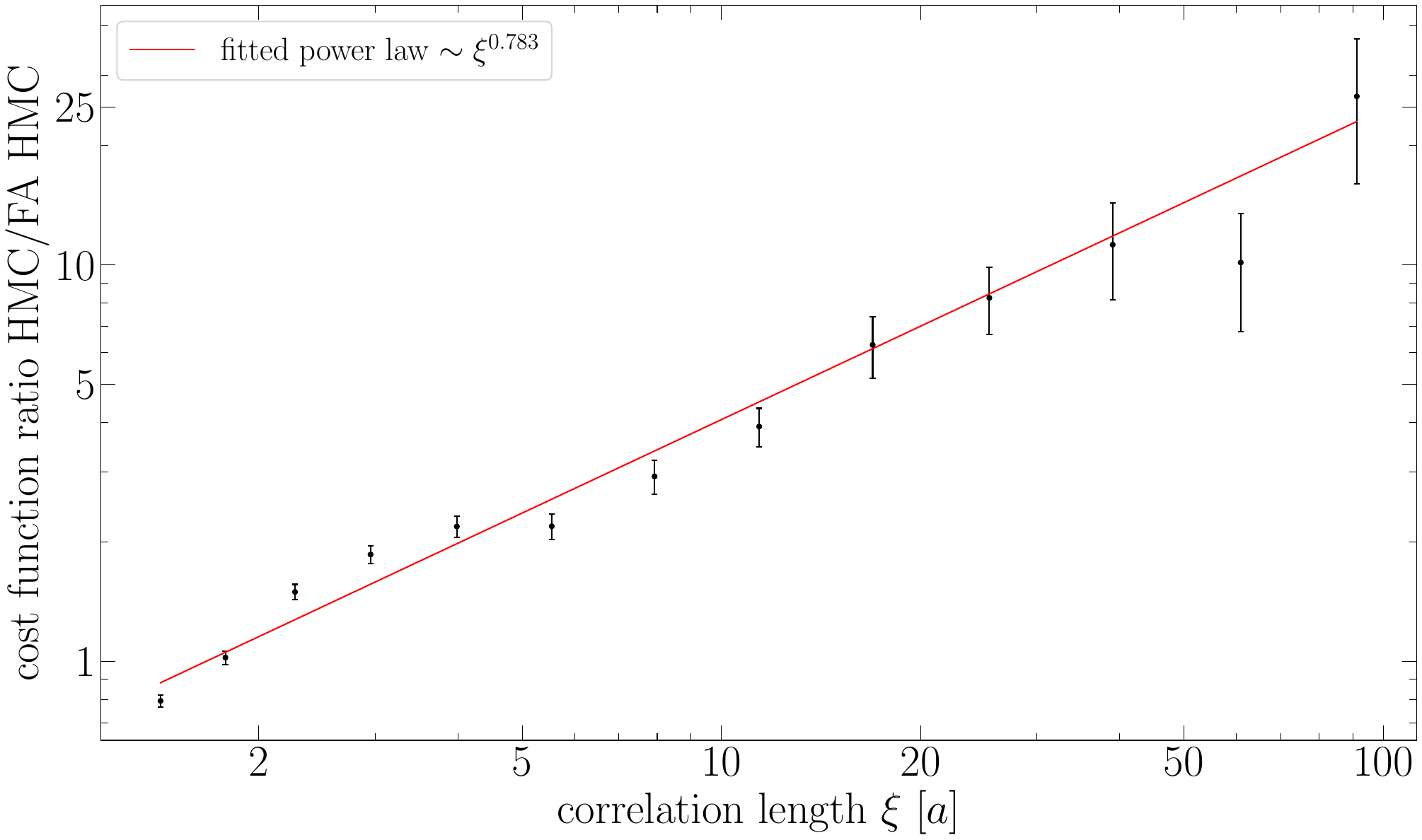}
    \caption{Ratio of the standard HMC and FA HMC cost functions against the system correlation length on a log-log scale with a fitted power law superimposed. The simulation parameters are identical to those used in Figure~\ref{fig:slowdown}.}
    \label{fig:cost_func}
\end{figure}

For FA HMC, we inferred the dynamical critical exponent $z$ to be $0.21 \pm 0.01$, offering a significant reduction compared to standard HMC with $z = 1.777 \pm 0.037$. Even though critical slowing down was not fully eliminated, employing Fourier Acceleration allows us to gain a factor of roughly $320$ in the susceptibility autocorrelation time at \mbox{$\beta=1.4,~\xi \approx 91.25$}. Consequently, one can expect that HMC requires $300$ times more trajectories than FA HMC to achieve comparable accuracy, translating to significantly larger CPU time requirements. In terms of the cost function, this corresponds to a factor of around $25$ as shown by plotting their ratio in Figure~\ref{fig:cost_func}. We find the cost advantage scales roughly as $\xi^{0.78}$. Indeed, as the functional dependency of the cost function is dominated by $\sqrt{\tau}$ and as the simulation time and acceptance rate are approximately the same for both algorithm choices, the same scaling is also found from half the difference of the gradients in Figure~\ref{fig:slowdown}.

\section{Conclusions}

The main conclusion from this study is the significant reduction of the severity of critical slowing down when introducing Fourier Acceleration to the Hybrid Monte Carlo simulations of the principal chiral model. The value of the acceleration mass parameter was shown to be not critical and the natural choice of setting it equal to the inverse of the correlation length yields close to optimal acceleration. The simplicity of this and the parallels between the principal chiral model and physical gauge theories such as QCD further encourages investigating Fourier Acceleration in the context of the latter.\\
In addition, we computed the mass over $\Lambda$ ratio to confirm the asymptotic scaling prediction and find our results in agreement with other SU($N$) $\times$ SU($N$) studies.\\
To achieve these objectives, we developed a simulation and analysis package which can serve as a foundation for further studies of SU($N$) $\times$ SU($N$) principal chiral models and is available in \cite{JW_repo}. It will be interesting to see how the performance of FA HMC compared to standard HMC develops against $N$. Moreover, we implemented an efficient approach to compute \mbox{(auto-)correlation} functions, resulting in a negligible time requirement for data analysis.\\
While in this article for simplicity we have tried to keep the acceptance approximately constant to be able to compare different runs, a further interesting avenue would be to investigate the influence of the mass parameter on the acceptance.

\section*{Acknowledgements}

RH is supported in part by the STFC grant ST/P000630/1. For the purpose of open access, the authors have applied a Creative Commons Attribution (CC BY) licence to any author accepted manuscript version arising from this submission.




\appendix
\section{Distribution of Momenta in two Dimensions}
\label{app:momentum_dist}
We define the $D$ dimensional Fourier transform on a cubic lattice through
\begin{equation}
    f_{\mathbf{x}} = \mathcal{F}^{-1} \left [ \tilde{f} \right] \equiv \frac{1}{L^D} \sum_{\mathbf{k}} \tilde{f}_\mathbf{k} e^{2 \pi i \frac{\mathbf{k} \cdot \mathbf{x}}{L^D}}
\label{eq:FT_def}
\end{equation}
and in the following denote the real and imaginary components of the momenta $\tilde{p}_\mathbf{k}$ by superscripts $R, I$ respectively. 
To assure that the configuration space momenta are real, $\tilde{p}_\mathbf{k}$ are constrained by Hermitian symmetry. When $k_1, k_2 \in \{0, \frac{L}{2} \}$, the exponential factors in Equation~\ref{eq:FT_def} become real, demanding
\begin{equation}
    \tilde{p}_{0,0}^I = \tilde{p}_{\frac{L}{2},0}^I = \tilde{p}_{0,\frac{L}{2}}^I = \tilde{p}_{\frac{L}{2},{\frac{L}{2}}}^I = 0.
\label{eq:constraint_red}
\end{equation}
If only one component takes either of the special values, we find with $k=1, \dots, \frac{L}{2}-1$ that
\begin{align}
    \tilde{p}_{L-k,0} &= \tilde{p}_{k,0}^*, \quad \tilde{p}_{0,L-k} = \tilde{p}_{0,k}^*, \nonumber \\
    \tilde{p}_{L-k,\frac{L}{2}} &= \tilde{p}_{k,\frac{L}{2}}^*, \quad  \tilde{p}_{\frac{L}{2},L-k} =  \tilde{p}_{\frac{L}{2},k}^*
\label{eq:constraint_pink}
\end{align}
as one exponential factor remains complex. For all other cases, namely $k_1, k_2 = 1, \dots, \frac{L}{2}-1$, hermiticity requires
\begin{equation}
    \tilde{p}_{k_1,k_2} = \tilde{p}_{L-k_1,L-k_2}^* \quad \text{and} \quad \tilde{p}_{L-k_1,k_2} = \tilde{p}_{k_1,L-k_2}^*.
\end{equation}
We now seek to construct a real-valued object $\tilde{\Pi}_{k_1, k_2}$ such that $\sum_{k_1,k_2} \vert \tilde{p}_{k_1,k_2} \vert^2 = \sum_{k_1,k_2} \tilde{\Pi}_{k_1,k_2}^2$. 
Defining $\tilde{\Pi}_{k_1, k_2}$ row-wise as
\begin{equation}
    \left ( \tilde{p}_{k_1,0}^R, \sqrt{2}\tilde{p}_{k_1,1}^R, \dots, \sqrt{2}\tilde{p}_{k_1,\frac{L}{2}-1}^R, \tilde{p}_{k_1,\frac{L}{2}}^R, \sqrt{2}\tilde{p}_{k_1,\frac{L}{2}-1}^I, \dots, \sqrt{2}\tilde{p}_{k_1,1}^I \right )
\label{eq:Pi}
\end{equation}
achieves this. The factors of $\sqrt{2}$ are required as due to Hermitian symmetry \mbox{$\sum_{k_2=1}^{\frac{L}{2}-1} |\tilde{p}_{k_1, k_2}|^2 = \sum_{k_2=\frac{L}{2}+1}^{L-1} |\tilde{p}_{k_1, k_2}|^2$}. Hence Equation~\ref{eq:mom_dist} follows.

\section{Correlations via Fourier Transforms}
\label{app:cor_via_FT}
Consider the two-point correlation function of a generic observable $O$ with observations $O_m$ and average $\langle O \rangle$. Defining $f_m \equiv O_m - \langle O \rangle$ and denoting the discrete convolution by $*$ shows that correlations can be written as a convolution in the observation chain index $m$:
\begin{equation}
    \Gamma(t) = \sum_m (O_m - \langle O \rangle) (O_{m+t} - \langle O \rangle) = \sum_m f_m f_{m+t} = f*f.
\end{equation}
The cross-convolution theorem then yields
\begin{equation}
    \Gamma(t) \sim \mathcal{F}^{-1} \left [ \left ( \mathcal{F}[f] \right )^2 \right ]
\end{equation}
which can be computed efficiently using FFTs. In this study, the typical number of elements to correlate is $\mathcal{O} \left( 10^5 \right)$ for which we measured the FFT approach to be six orders of magnitude faster than the naive summation. However, care must be taken when using FFTs as their product yields the circular convolution of the original signals rather than the linear one. For physical correlations, the former automatically accounts for the periodic lattice boundary conditions and is therefore desirable. On the other hand, for autocorrelations in the chain of measurements, a linear convolution is required. This can be obtained by simply padding the original data with sufficiently many zeros such that the circular mixing with these leaves the original data unchanged. The padding is usually done in powers of two as FFT routines work most efficiently with such a number of elements. Once all Fourier transforms have been taken, one must truncate the introduced pads to get the final autocovariance function result.\\
To employ this approach for the computation of the wall-to-wall correlation in Equation~\ref{eq:wallwallcor}, two observations are required. First, for $A,B \in \text{SU}(2)$ and $C=AB$ with respective parameter vectors $a,b,c$, we have $\tr C = 2c_0 = 2(a_0b_0 - a_jb_j) \in \mathbb{R}$. Further, based on Equation~\ref{eq:phi_parameterisation}, we find that Hermitian conjugation leaves the $0$th component of the parameter vector invariant while the spatial components are negated. Hence
\begin{align}
    \C_{ww}(d) &= \left \langle \frac{4}{PL^2} \sum_{i} \left \{ \sum_{v,w} (\phi_m(v,i))_0(\phi_m(w,i+d))_0 \right. \right. \nonumber \\  
    &+ \left. \left. (\phi_m(v,i))_j(\phi_m(w,i+d))_j \vphantom{\sum_{v,w}} \right \} \vphantom{\sum_i} \right \rangle_m \nonumber \\
    &= \left \langle \frac{4}{PL^2} \sum_{i} (\Phi_m(i))_{\alpha}(\Phi_m(i+d))_{\alpha} \right \rangle_m  \label{eq:wwcor_fast}
\end{align}
with $\Phi_m(i) = \sum_p \phi_m(v,i)$ and where the summation over $\alpha=0,1,2,3$ in Euclidean space is implied. Note that for each value of $\alpha$, the remaining sum is a convolution in the variable $i$ of the scalar quantity $\Phi_m$ and is thus efficiently computed using FFTs. Similarly, we write Equation~\ref{eq:def_chi} as
\begin{align}
    \chi &= \frac{2}{L^2} \sum_j \sum_i \left\{ \sum_{v,w} \phi(v,i)_{\alpha} \phi(w,j)_{\alpha} \right\} \nonumber \\
    &= \frac{2}{L^2} \sum_j \left\{ \sum_i (\Phi(i))_{\alpha} (\Phi(j))_{\alpha} \right\}.
\label{eq:chi_crosscor}
\end{align}
Due to the lattice periodicity and since $i$ and $j$ both index columns of the lattice, one can always express $i$ as a shifted version of $j$ such that the inner bracket is simply the convolution already encountered in Equation~\ref{eq:wwcor_fast}.



\end{document}